# Tailoring ink-substrate interactions via thin polymeric layers for high-resolution printing

*Aleksander Matavž,\*,[†,⊥] Vid Bobnar,[†,⊥] and Barbara Malič [†,⊥]*

† Jožef Stefan Institute, Jamova cesta 39, 1000 Ljubljana, Slovenia

⊥ Jožef Stefan International Postgraduate School, Jamova cesta 39, 1000 Ljubljana, Slovenia

\* Corresponding Author: aleksander.matavz@ijs.si

**ABSTRACT**

The surface properties of a substrate are among the most important parameters in the printing technology of functional materials, determining not only the printing resolution but also the stability of the printed features. This paper addresses the wetting difficulties encountered during inkjet printing on homogeneous substrates as a result of improper surface properties. We show that the wetting of a substrate and, consequently, the quality of the printed pattern, can be mediated through the deposition of polymeric layers that are a few nanometers thick. The chemical nature of the polymers determines the surface energy and polarity of the thin layer. Some applications, however, require a rigorous adjustment of the surface properties. We propose a simple and precise method of surface-energy tailoring based on the thermal decomposition of poly(methyl methacrylate) (PMMA) layers. A smooth transition in the wetting occurs when the thickness of the PMMA layer approaches zero, probably due to percolating the underlying surface of the substrate, which enables the inkjet printing of complex structures with a high resolution. In particular, the wetting of three substrate–ink systems was successfully adjusted using the thin polymeric layer: (i) a tantalum-oxide-based ink on indium-tin-oxide-coated glass, (ii) a ferroelectric lead zirconate titanate ink on a platinized silicon substrate, and (iii) a silver nanoparticle ink on an alumina substrate.

**Introduction**

The wetting of a solid surface is one of the major concerns in large-scale industrial processes such as printing, painting, gluing and cleaning. It can be quantified in terms of the contact angle, $\theta$, which is the angle between the liquid–vapor interface and the surface of the substrate.[1] In printing applications the resolution and the stability of printed features depend on the contact angle.[2] For example, small contact angles result in over-spilling of the ink and limit the patterning, while large contact angles result in the formation of instabilities that prevent the stable coalescence of a drop. For $\theta$ between 0° and 90°, the printing resolution depends on the size of the liquid drop on the selected substrate. Small drops—for such drops the shape deformations due to gravitational forces are negligible—adopt the shape of a spherical cap with a radius of

$$r = \sqrt[3]{\frac{3\,V_d\,sin^3\theta}{\pi(2 - 3cos\theta + cos^3\theta)}}, \qquad (1)$$

where $V_d$ is the volume of the drop.[3] This relation shows that a relatively small reduction in the contact angle from 30° to 10° increases the radius of the printed drop by nearly 50%. It is self-implied that the resolution improves for ink–substrate combinations with larger contact angles, but larger contact angles also promote the formation of instabilities, such as bulging of the printed lines.[2,4] Precise control over the wetting of a substrate by a particular ink is therefore essential in printing applications.

The tailoring of wetting on a local scale has received a lot of attention as it could enable the fabrication of self-aligned structures and channels with a sub-micrometer precision.[5–7] On the other hand, the manipulation of wetting on the macroscopic level and its influence on the printed features has not been studied extensively. van Osch et al. demonstrated that the width of a printed line depends on the surface free energy (SFE) of the substrate; printing on polymeric substrates with low or high SFEs results in narrow lines with bulges, or broad and continuous lines, respectively.[8] The authors used polymeric substrates with different SFEs; although in practice it would be more practical to modify only the surface of the substrate. Several approaches have been used to impact the surface properties of different materials, including cleaning,[9] plasma and corona treatment,[10–12] UV/O$_3$ exposure,[13] depositing self-assembled monolayers,[14] and chemical treatment.[15] The efficiency of these approaches, however, depends strongly on the chemical nature of the substrate. Moreover, plasma and UV/O$_3$ treatments require special equipment that is not always available in clean-room facilities.

In this paper we describe the modification of surface properties for various substrates using thin polymeric layers and show how the surface properties relate to the morphology of inkjet-printed structures. The surface properties of poly(styrene), poly(methyl methacrylate), poly(vinyl alcohol), and poly(vinyl pyrrolidone) layers are assessed using the Owens–Wendt–Rabel–Kaelble (OWRK) model, which is discussed in detail in the next section. The inkjet printing of two inks with distinct polarities is used to connect the printing outcome with the SFE of the polymeric layers. The few nanometers thick layers functions only to regulate the substrate–ink interface (to influence the wetting), and could be afterwards removed from the substrate. This concept proved sufficient only for some ink–substrate combinations and pattern geometries—other combinations required a more precise adjustment of the wetting. We introduce a thermal decomposition of the poly(methyl methacrylate) layer on glass substrates as a method for precisely adjusting the surface properties, which is used to optimize the print quality and the resolution. In the last section we provide some practical examples for the wetting adjustment of functional-oxide inks on different substrates, such as indium-tin-oxide-coated glass, platinized silicon, and alumina.

**Owens–Wendt–Rabel–Kaelble (OWRK) model**

The SFE of a solid cannot be directly measured, but is estimated using models. The Owens–Wendt–Rabel-Kaelble (OWRK) model is a two-component model based on the Fowkes assumption that two types of interactions exist at the solid–liquid interface.[16] The SFE of a solid or a liquid is divided into its dispersive and polar components, which are related by geometrical mean values. The solid–liquid interfacial tension, $\gamma_{SL}$ is

$$\gamma_{SL} = \gamma_L + \gamma_S - 2\left(\sqrt{\gamma_L^D \gamma_S^D} + \sqrt{\gamma_L^P \gamma_S^P}\right), \tag{2}$$

where $\gamma_L$ is the overall SFE of the liquid, $\gamma_S$ is the overall SFE of the solid, and the $D$ and $P$ superscripts correspond to the dispersive and polar contributions to the SFE. Inserting the above relation into Young's equation, $\gamma_S = \gamma_{SL} + \gamma_L \cos(\theta)$, results in

$$\frac{\gamma_L(\cos(\theta) + 1)}{2\sqrt{\gamma_L^D}} = \sqrt{\gamma_S^P} \cdot \frac{\sqrt{\gamma_L^P}}{\sqrt{\gamma_L^D}} + \sqrt{\gamma_S^D}, \tag{3}$$

which is actually a linear equation. In a polar-coordinate system, Equation 3 becomes

$$r(\varphi) = \left(\frac{\sqrt{\cos(\varphi) \cdot \gamma_S^D} + \sqrt{\sin(\varphi) \cdot \gamma_S^P}}{\cos(\varphi) + \sin(\varphi)}\right)^2 \left(\frac{2}{1 + \cos(\theta)}\right)^2. \tag{4}$$

The SFE parameters of a saurface are calculated by measuring the contact angles of at least two probe liquids with known $\gamma_L$, $\gamma_L^D$ and $\gamma_L^P$, and then performing a linear fit in the $\gamma_L(\cos(\theta) + 1)/2\sqrt{\gamma_L^D}$ vs. $\sqrt{\gamma_L^P}/\sqrt{\gamma_L^D}$ plot. The same set of probe liquids should be used when comparing the surface properties of different solids to improve the reliability of the method.[17] The probe liquids selected for our study, together with their surface-tension values, are collected in **Table 1**.

OWRK model only roughly estimates the surface energy of a solid and the evaluation relies on certain mathematical assumptions. Furthermore, the estimation of solid's SFE is also obstructed by the fact that Young contact angle cannot be directly measured—contact angle measurements access only the apparent contact angle.[18] The methods for the evaluation of Young contact angle are still under intensive debate and research.[19,20] Yet, the simplicity and relatively good agreement with the experiments makes the OWRK model the most widely used SFE model. For a substrate with given $\gamma_S^D$ and $\gamma_S^P$, the OWRK model allows a convenient prediction of the wetting-ability for a random liquid with given $\gamma_L^D$ and $\gamma_L^P$ via the construction of wetting envelopes. An example of a wetting-envelope diagram is shown in **Figure 1**. The solid in this example has $\gamma_S^D = 20$ mJ/m² and $\gamma_S^P = 20$ mJ/m². Different wetting situations are exemplified by the liquids A, B and C, which all have different polar and dispersive components of the surface tension. It is important to note that—according to the OWRK model (and other multi-component models)—the wetting does not directly depend on the overall SFE, but is determined by the degree of polar–polar and dispersive–dispersive interaction between the solid and the wetting liquid.

**Table 1.** Set of probe liquids with corresponding surface-tension values (from Ref. 17).

| Probe liquid | $\gamma_L$ (mJ/m²) | $\gamma_L^D$ (mJ/m²) | $\gamma_L^P$ (mJ/m²) |
|---|---|---|---|
| Water | 72.8 | 26.4 | 46.4 |
| Glycerol | 63.4 | 37.0 | 26.4 |
| Benzyl alcohol | 39.0 | 30.3 | 8.7 |
| Toluene | 28.4 | 26.1 | 2.3 |

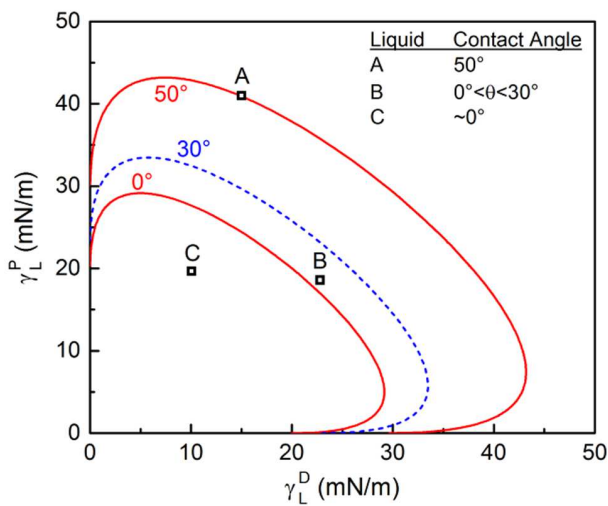

**Figure 1.** Wetting envelopes of a solid ($\gamma_S = 40$ mJ/m$^2$, $\gamma_S^D = 20$ mJ/m$^2$, $\gamma_S^P = 20$ mJ/m$^2$) constructed for θ of 0°, 30° and 50°. The positions of $\gamma_L^D$–$\gamma_L^P$ points predict $\theta = 50°$ for the liquid A, $0° < \theta < 30°$ for the liquid B, and $\theta \approx 0°$ for the liquid C.

**Experimental**

**Materials**

Poly(methyl methacrylate)–PMMA (500 kDa), poly(vinyl alcohol)–PVA (98–99% hydrolyzed, 57–66 kDa), and poly(vinyl pyrrolidone)–PVP (40 kDa) were purchased from Alfa Aesar. Poly(styrene)–PS (500 kDa) was synthesized by anionic polymerization at the National Institute of Chemistry, Slovenia. 0.5 wt% polymer solutions were agitated until complete dissolution of the polymers in selected solvents: PMMA in anisole (99%, Alfa Aesar), PS in toluene (99.5%, Carlo Erba Reagents), PVA in deionized water, PVP in 2-methoxyethanol (99.3+%, Alfa Aesar). The glass substrates were Corning XG Eagle. The liquids used for the surface-energy evaluation were water (Milli-Q, $\rho = 18.2$ MΩ/cm), glycerol (99.95%, Alfa Aesar), benzyl alcohol (99+%, Alfa Aesar) and toluene (99.5%, Carlo Erba Reagents).

Two inks were formulated for the inkjet printing trials. The first ink, denoted as MOE-PD, was prepared by dissolving 2890 mg of $In(NO_3)_3$ (99.99%, Alfa Aesar) and 295 mg of $Zn(NO_3)_2$ (99+%, Sigma Aldrich) in 20 mL of a solvent mixture of 2-methoxyethanol (99.3+%, Alfa Aesar) and 1,3-propanediol (98%, Sigma Aldrich) in 45:55 volume ratio. Finally, 1.14 mL of acetic acid (99.8+%, Sigma Aldrich) was added. The second ink, denoted as OCT ink, was prepared by dissolving 810 mg of tantalum ethoxide (99.99%, H. C. Starck) and 420 mg of diethanolamine (99%, Alfa Aesar) in 20 mL of 1-octanol (99%, Merck).

Additional inks were formulated to assess the wetting of indium-tin-oxide-coated glass, platinized silicon wafer, and alumina. The tantalum-oxide-based ink (denoted as TAS ink) consisted of tantalum ethoxide (H.C. Starck, 99.99%), aluminum *sec*-butoxide (Sigma Aldrich, 97%) and silicon ethoxide (Alfa Aesar, 99,9%) dissolved in a solvent mixture of 2-methoxyethanol, glycerol and 1,3-propanediol in 65:25:10 volume ratio. The details of the ink preparation can be found in Ref 21,22. To prepare the $Pb(Zr_{0.53}Ti_{0.47})O_3$ (PZT) ink, lead acetate (99–103%, Alfa Aesar), zirconium *n*-butoxide (80 wt% in 1-butanol, Alfa Aesar) and titanium *n*-butoxide (99+%, Alfa Aesar) were first dissolved and reacted in 2-methoxyethanol, analogous to the synthetic route described in Ref 23. Then, 55 vol% of 1,3-propanediol was added to yield the PZT ink. The Ag ink was a commercially available silver dispersion ink, Suntronic U5714. The ink contains ethanol (~40 vol%) and ethylene glycol (~40 vol%) as the main solvents.[24]

**Preparation of thin polymeric layers**

Glass substrates were cleaned by wiping them with a cloth soaked in an aqueous detergent solution, followed by a sequential ultrasonic cleaning in a detergent solution and deionized water. The substrates were dried by blowing with nitrogen gas and heating at 350 °C for 10 minutes. The polymer solutions were filtered through a 0.2-µm PTFE filter prior to spin coating. The layers were prepared by completely covering the substrate with a polymer solution and spin-coating at 3000 rpm for 30 s. Note that two different thicknesses of the PMMA layer were obtained by spin coating either the 0.5 wt% or 3 wt% PMMA solution. The samples were dried on a hotplate at 200 °C for 10 minutes immediately after spin coating. Selected PMMA layers were heated on a hotplate at 350 °C in air.

**Inkjet printing**

The inkjet printing was performed under ambient conditions using a Dimatix DMP 2831 piezoelectric printer equipped with a 10-pL cartridge (DMC-11610). The MOE-PD and OCT inks were printed at a jetting frequency of 10 kHz with a distance between the cartridge and the substrate of 0.7 mm and a drop spacing of 20 μm. The temperature of the printer platen and nozzles was set to ambient (22–23 °C). The printing pattern consisted of an individual line (3 mm in length) and a 500×500 μm$^2$ square.

**Characterization of polymeric layers**

The contact angle, $\theta$, was measured under ambient conditions, $T = 22\pm1$°C, $RH = 30\pm5$, by a sessile-drop method using a Krüss DSA20E tenziometer. The contact angle of the as-placed drops was extracted from the drop's side-view image by Young–Laplace fit for $\theta > 10°$, or by circle-segment fit for $\theta < 10°$. The advancing and receding contact angles were determined by pumping the liquid into or out of the drop while recording the drop from a side-view and performing a tangent fit at the solid–liquid–vapor interface. The surface free energy and wetting envelopes were calculated using the OWRK model. The surface tension of the liquids was measured by a pendant-drop method. The polar and dispersive components of the surface tension of the inks were calculated by measuring the contact angle of a selected ink on PTFE and assuming purely dispersive character of its surface ($\gamma_S^D = 18$ mJ/m$^2$, $\gamma_S^P = 0$ mJ/m$^2$). The surface morphology was studied using an Asylum Research MFP-3D-STM atomic force microscope (AFM) in tapping mode using Olympus OMCL-AC240TS-R3 cantilevers. The thickness of the polymeric layers was estimated by producing a scratch with a sharp needle, recording the thickness profile across the scratch with AFM, and subtracting the height differences. The thicknesses of the thin polymeric layers deposited from the 0.5 wt% solution were: 3 nm for PVA, 5 nm for PVP, 10 nm for PMMA, 23 nm for PS.

# Results

## Glass and thin polymeric layers

Contact-angle measurements of the probe liquids provided the data for the calculation of the surface free energy (SFE) of the investigated surfaces. Uncoated glass substrates exhibit a large SFE and a high surface polarity (see **Table 2**). According to the OWRK model, most of the solvents (except those that are highly nonpolar) would wet the glass substrate completely. Such wetting behavior is desirable in some applications; however, it represents a large impediment to the patterning of well-defined structures using inkjet printing.

Coating a glass substrate with a thin polymeric layer affected its SFE, which depends on the chemical nature of the polymer, while the surface morphology (and surface roughness) remained unaffected. PVA and PVP exhibit a hydrophilic character and the layers accordingly exhibited a high surface polarity of about 90%, as well as a high overall SFE (**Table 2**). In contrast, PMMA and PS are hydrophobic, thus the respective layers exhibit a low surface polarity and a significantly reduced overall SFE when compared to the bare glass. The zero polarity of a PS-coated substrate layer implies that such a surface is incapable of forming any polar interactions with the liquids used in this study.

Large differences in the surface properties of glass and polymeric layers are expected to significantly impact on the wetting behavior of a particular liquid. To assess these differences, we formulated two inks—a moderately polar MOE-PD ink and a non-polar OCT ink (see **Table 2**)—and compared the printing outcome to the wetting diagrams. The results for the PVA layers are not reported here as they exhibited nearly identical surface properties to glass.

The position of the $\gamma_L^D$–$\gamma_L^P$ point of the MOE-PD ink inside the 0° wetting envelopes of the glass and the PVP layer implies that the ink wets these surfaces completely, which is reflected in the large width of the printed lines (first row, **Figure 2**) and has been also confirmed by the contact angle measurements ($\theta \approx 0°$).

**Table 2.** Surface free energy of bare glass substrate, glass coated with a thin polymeric layer and inks used in printing trials.

|  | Sample | $\gamma_S, \gamma_L$ (mJ/m²) | $\gamma_S^D, \gamma_L^D$ (mJ/m²) | $\gamma_S^P, \gamma_L^P$ (mJ/m²) | Polarity (%) |
|---|---|---|---|---|---|
| **Substrate** | Bare glass | 82 | 7 | 75 | 91 |
| **Hydrophilic polymers** | Glass/PVA | 78 | 6 | 72 | 92 |
|  | Glass/PVP | 75 | 9 | 66 | 88 |
| **Hydrophobic polymers** | Glass/PMMA | 33 | 26 | 8 | 23 |
|  | Glass/PS | 36 | 36 | 0 | 0 |
| **Inks** | MOE-PD | 38.7 | 27.8 | 10.9 | 28 |
|  | OCT | 26.9 | 25.2 | 1.7 | 6.3 |

According to the wetting diagram the same ink should only partially wet the PMMA and PS layers. This correlates well with the printing results, where a small width of the printed lines corresponds to a large contact angle.[2] The advancing contact angles measured by a modified sessile drop method were 37° and 60° for PMMA and PS layers, respectively, which is in accordance with the predicted values from the wetting diagrams. In the case of PS-coated glass, continuous lines could not be printed; instead, the lines broke into individual large drops (bottom-right part of **Figure 2**). It is known that stable and continuous lines can be printed only when the receding contact angle is zero, or close to zero if there is a large contact angle hysteresis.[25] In the case of non-zero receding contact angle— and hence, unfixed contact line—the liquid bead breaks into separate drops due to the Rayleigh instability. As the receding contact angle of MOE-PD ink on PS was 56°, this is probably the reason why stable lines did not form on those samples. All the other samples exhibited ~0° receding contact angle.

Due to the lower OCT ink polarity, the predicted wetting behavior is different from that of the MOE-PD ink. The printing of the OCT ink on glass produced narrow lines, which is in accordance with the large contact angle predicted by the wetting diagram (inset of **Figure 2**, glass). Due to the slightly stronger dispersive character of the PVP, printing the OCT ink yielded broader lines than for the bare glass. The measured advancing contact angles on glass and PVP substrates were 35° and 15°, which also fits well with the predictions of the OWRK model. For both the PMMA and PS layers the $\gamma_L^D$–$\gamma_L^P$ of the OCT ink lies inside the 0° wetting envelopes, and accordingly, the printing of the OCT ink produced wide lines (bottom row of **Figure 2**).

The results presented above clearly demonstrate that two-component model of SFE is required to explain the wetting of the inks. For example, SFE of glass is much higher than those of PMMA layer (see **Table 2**), however, the OCT ink wets the glass better. In this case, a direct comparison of overall SFE leads to a wrong prediction of the wetting. The same occurs when comparing the wetting of OCT and MOE-PD ink on the glass, whereas the MOE-PD ink has larger SFE but exhibits a lower contact angle than OCT ink. The presence of a solute and additives, such as acetic acid in the OCT ink or diethanolamine in the MOE-PD ink, could also affect the wetting behavior. More sophisticated models, e.g., the van Oss-Good-Chaudhury model, could be used to distinguish between the Lewis acid-base contributions; however, a higher complexity of the model makes the analysis in terms of wetting diagrams impossible. Even though OWRK model provides only approximation of the SFE, a good correlation to the experimental results was found for all investigated combinations.

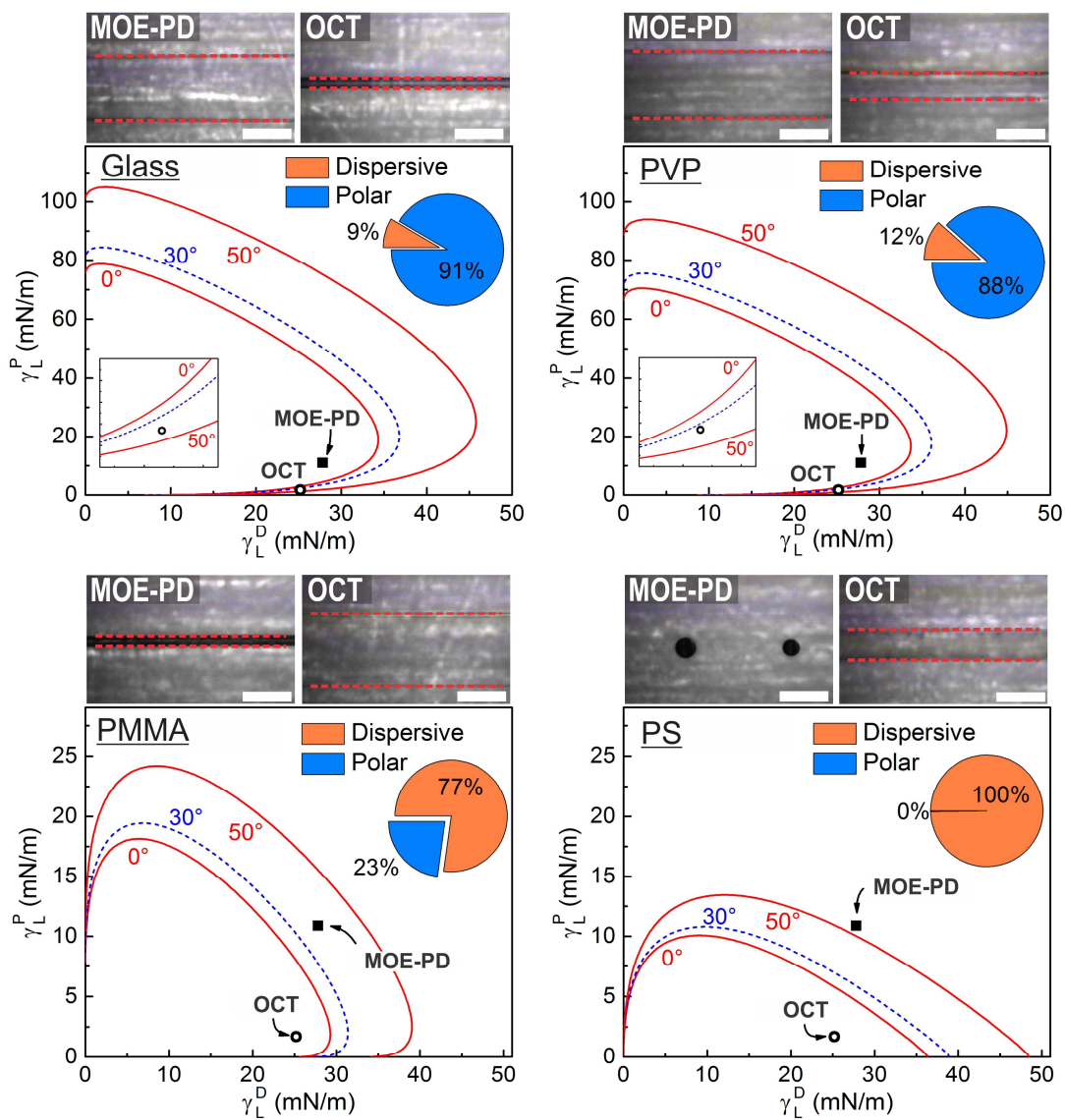

**Figure 2.** Wetting envelopes represented as $\gamma_L^D$–$\gamma_L^P$ plots (designed with two different y-axes scales) and optical micrographs of printed lines of two ink formulations on glass, and PVP-, PMMA-, or PS-coated glass. Scale bars in optical micrographs correspond to 200 μm. Pie charts show the fractions of polar and dispersive contributions to the SFE of glass and polymeric layers. The red dotted lines in the optical micrographs mark the edges of the printed lines; however, in the case of PS-coated glass, the contact angle of the MOE-PD was too large to print a continuous line.

The ability to deposit well-defined patterns is a prerequisite for the realization of electronic devices by printing technology. Patterns often consist of different geometrical shapes, such as dots, lines, and rectangles. While the drops are always stable, the lines and the squares may exhibit instabilities. The bulging of lines is promoted by a large contact angle, low printing speed and small drop spacing, while breaking of lines appears due to Rayleigh instability caused by a non-zero receding contact angle.[4] Similarly to lines, rectangular patterns also exhibit instabilities, such as the breaking of continuous structures into separate droplets and rounding of the corners, which are often related to a non-zero receding contact angle.[26,27]

Our experiments show that the stability of 2-dimensional patterns, *i.e.*, 500×500 μm$^2$ squares, is even more sensitive to the surface properties than the stability of the lines. Even though the lines were continuous when printing the OCT ink on glass (or similarly the MOE-PD on PMMA, see **Figure 2**), the continuity of the square patterns broke for the same ink–substrate combinations (**Figure 3**). On the other hand, the ink–substrate combinations that produced wide lines formed undefined shapes when printing a square pattern, *e.g.,* MOE-PD on the glass. The deterioration of the square shape is a consequence of the ink overspreading driven by a small contact angle. The overspreading could, to some degree, be minimized by increasing the drop spacing, *e.g.,* in the cases of printing OCT on PVP and PS layers. However, for the MOE-PD ink on the glass, PVA or PVP layers, as well as the OCT ink on the PMMA layer, the over-spreading is too extensive and inhibits the deposition of well-defined patterns.

The results clearly reveal that the ink–substrate interactions are a key factor in high-quality printing. The wetting behavior of the thin polymeric layers on a glass substrate is determined by the intrinsic properties of the polymer. However, printing geometrically more complex objects, such as those shown in **Figure 3**, requires a precise adjustment of the substrate surface properties.

In the following section we describe a simple, but effective, approach to the precise modification of the surface properties by the thermal decomposition of the PMMA layers.

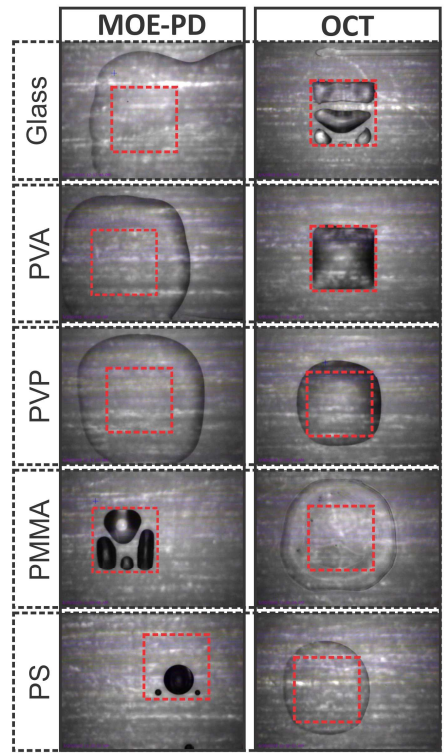

**Figure 3.** Printed square pattern of MOE–PD and OCT inks on a bare glass and glass substrates coated with a thin polymeric layer. The user-defined 500×500 μm² square pattern is marked in red.

**Thermal decomposition of PMMA layers**

Upon heating PMMA degrades almost completely into its monomeric units.[28] The decomposition temperature is strongly affected by the atmosphere in which the degradation occurs as well as by the polymerization method and the molecular weight of the polymer.[29] In air, the onset of the thermal decomposition of the PMMA powder is typically at ~250 °C.[30] The decomposition of thin layers, however, is expected to differ somewhat from the decomposition of powder.[31] Isothermal heating at 350 °C of a PMMA layer on glass resulted in a steady decrease of its thickness (**Figure**

**4**) due to evaporation of the decomposition products and densification of the layer. The thickness of the initially 175-nm-thick layer decreases to zero after 20 minutes of heating at 350 °C, indicating a complete decomposition of the polymer.

The impact of thermal decomposition of the PMMA layer on its surface properties was evaluated in the following way. The advancing contact angle of MOE-PD ink was measured on a PMMA layer that was heated at 350 °C for different times and afterwards cooled to room temperature. **Figure 4** presents the contact angles of the MOE-PD ink on these substrates. For heating times up to about 15 minutes the contact angle remains almost unchanged at about 35°. However, when the PMMA layer thickness approaches zero (this region is indicated by the vertical dashed lines in **Figure 4**) we observe a rapid reduction in the contact angle of the MOE-PD ink.

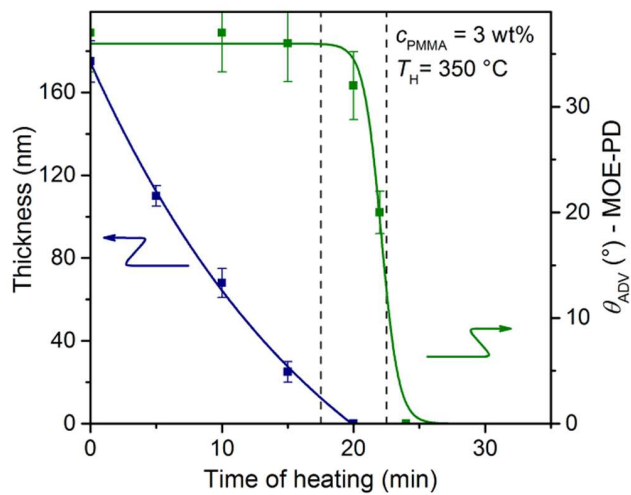

**Figure 4.** Thermal decomposition of PMMA layer at 350 °C in air. The layer was deposited from a 3 wt% PMMA solution. The plot shows the thickness of the layer and the advancing contact angle of the MOE–PD ink as a function of the heating time.

To provide a better insight into the origins of the observed wetting transition we performed a surface-energy analysis of the PMMA layers heated for different times at 350 °C. Note that in these experiments the polymer concentration was 0.5 wt%, which strongly reduced the initial layer thickness to about 10 nm. Although the same experiment could be performed with thicker layers, it is easier and more elegant to control the wetting transition of the thinner layers. Due to a lower thickness, the layer decomposes at shorter heating times. As shown in **Figure 5**a, polar and dispersive contributions to the surface energy of the initial 10-nm-thick and 1-minute heated PMMA layer are very similar, while 4 minutes of heating already initiates the wetting transition. For $t \geq 4$ minutes, the polar component of the SFE strongly increases and, simultaneously, the dispersive component decreases. After 20 minutes of heating, the SFE becomes similar to that of bare glass. The evolution of the wetting envelopes with the heating time of the PMMA layer is shown in **Figure 5**b. As expected from the surface-energy analysis, the wetting envelopes drastically change with the degree of the layer decomposition. The most pronounced change in wetting envelopes is the upward stretch due to a strong increase in the polar contributions with the heating time. This stretching is accompanied by a slight shift of the wetting envelopes to the left—a consequence of the dispersive component reduction. The left-shift can be clearly observed when comparing the layers heated for 0 and 4 minutes. The inset of **Figure 5**b shows that additional heating shifts the wetting envelopes even more to left. Both the upward stretching and left-shift of the wetting envelopes have a large impact on the wetting behavior, which will be discussed in the following paragraphs.

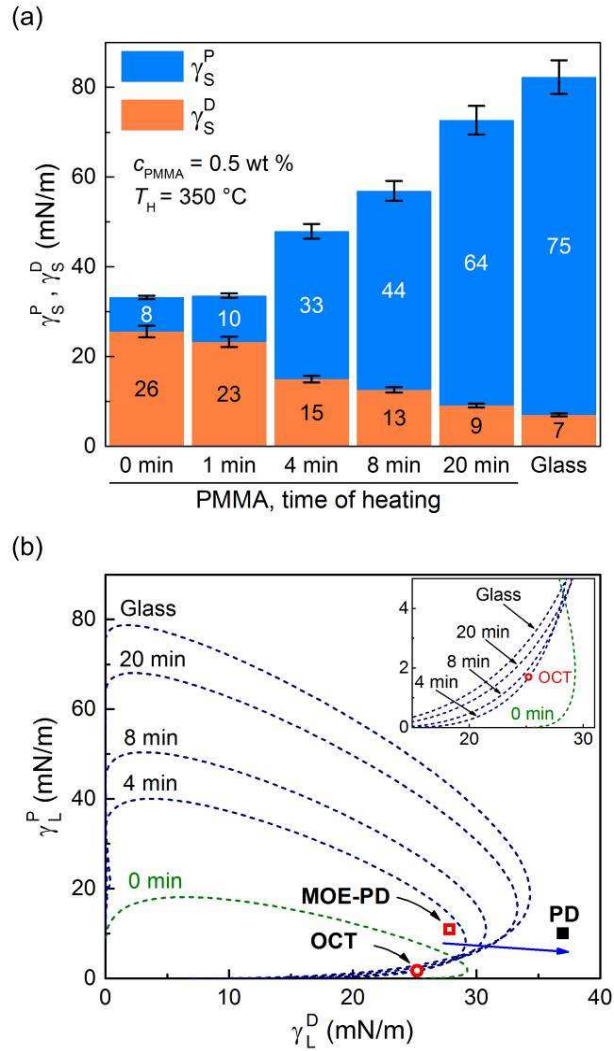

**Figure 5**. (a) Polar and dispersive contributions to surface free energy of the 10-nm-thick PMMA layers that were heated for different times at 350 °C. The layer was deposited from a 0.5 wt% PMMA solution. (b) Wetting envelopes calculated for 0° contact angle. The wetting envelope of sample heated for 1 minute is similar to that of the initial PMMA layer and is not shown. The inset shows the evolution of wetting envelopes near the $\gamma_L^D$–axis. In both plots the data for the glass substrate are included for comparison. The bottom plot also includes the $\gamma_L^D$–$\gamma_L^P$ points of MOE-PD, OCT and 1,3-propanediol (PD), and the change of the surface energy due to solvent evaporation of the MOE-PD is indicated by a blue arrow towards the point PD.

The transition of the surface properties upon heating—from PMMA-like toward glass-like—can be explained by a situation in which the glass substrate percolates in random small spots through a decomposing PMMA layer. In such spots the surface properties correspond to the ones of the glass substrate, while the total area of the exposed spots determines the macroscopic properties. We thus express the surface concentration of polymer molecules on the substrate as the surface coverage, $n$, which is the area covered by polymer molecules divided by the total area. For $n = 1$, the surface is completely covered by the polymer molecules and the surface properties correspond to those of the polymer (exemplified by Stage 1 in **Figure 6**). For $n = 0$, there are no polymer molecules at the surface–air interface and the properties correspond to those of the substrate, see Stage 3, **Figure 6**. For $0 < n < 1$, the properties strongly depend on the value of $n$. The polar and dispersive components relate to the surface coverage with

$$\gamma^P(n) = \left(\gamma^P_{polymer} - \gamma^P_{substrate}\right) \cdot n + \gamma^P_{substrate} \qquad (5)$$

$$\gamma^D(n) = \left(\gamma^D_{polymer} - \gamma^D_{substrate}\right) \cdot n + \gamma^D_{substrate} \qquad (6)$$

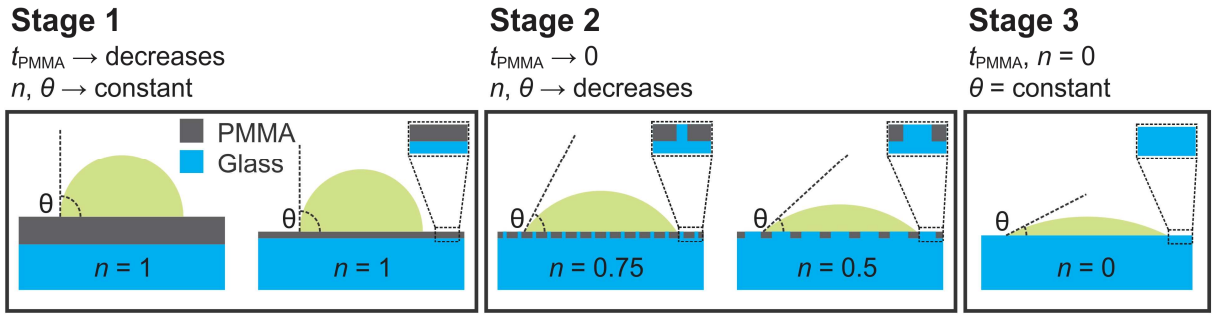

**Figure 6**. Schematic presentation of the wetting transition for the PMMA layer depending on its coverage ($n$) of the glass substrate from $n = 1$ in Stage 1 to $n = 0$ in Stage 3. The contact angle of the liquid (marked in green) on the PMMA layer is much higher than on the glass. The drop,

PMMA layer and glass substrate are not in scale for a better visualization of the decomposition process.

Equations 5 and 6 are valid in the case of non-interacting percolative spots that are much smaller than the liquid feature wetting the substrate. Inserting the boundary conditions ($n = 1$ or $0$) into the equations shows a clear transition from polymer-like toward substrate-like surface properties. The situation is schematically represented by Stage 2 in **Figure 6**.

Printing a square pattern on the PMMA layers after different stages of the thermal decomposition assessed the viability of the approach. Printing on the as-deposited PMMA layers results in a discontinuous layer of the MOE-PD and over-spreading of the OCT ink, as already discussed in the previous section. Isothermal heating of the layer for 4 minutes changed the surface properties and also the morphology of the square pattern printed from the MOE-PD ink. The poorly defined edges of the printed square (left column in **Figure 7**) suggest that further heating is required to create optimal patterning conditions. Upon heating for a total of 8 minutes the deposition of well-defined structures of MOE-PD ink was enabled, whereas longer heating times decreased the printing resolution and resulted in over-spilling of the ink. The printing results (**Figure 7**) show some inconsistency with the wetting envelopes presented in **Figure 5**b. Although the position of the MOE-PD $\gamma_L^D$–$\gamma_L^P$ point in the wetting envelope of the layer heated for 4 and 8 minutes indicates that the ink should completely wet the substrate, the printing resulted in a partial wetting. This inconstancy is probably related to solvent evaporation during the printing process, which induces a shift from the MOE-PD toward the PD $\gamma_L^D$–$\gamma_L^P$ point in the wetting diagram (indicated by a blue arrow in **Figure 5**b)—a consequence of such a shift is a larger actual contact angle for the printed ink.[24]

A different wetting situation was observed in the case of the OCT ink. A 4-minute heating at 350 °C already resulted in optimal surface properties for the layer and enabled patterning of well-defined squares (right column in **Figure 7**). Any additional heating of the PMMA layer resulted in a degradation of the print quality caused by a too large contact angle. The printing results are in good correlation with the wetting envelopes presented in **Figure 5**b, whereas a right shift of the wetting envelopes with the heating time indicates an increase in the contact angle.

The results presented above show that heating the PMMA layer at 350 °C produces a smooth transition from a large to a small contact angle for the MOE-PD ink (and vice versa for the OCT ink), which can be exploited for a precise tailoring of the wetting of glass substrates.

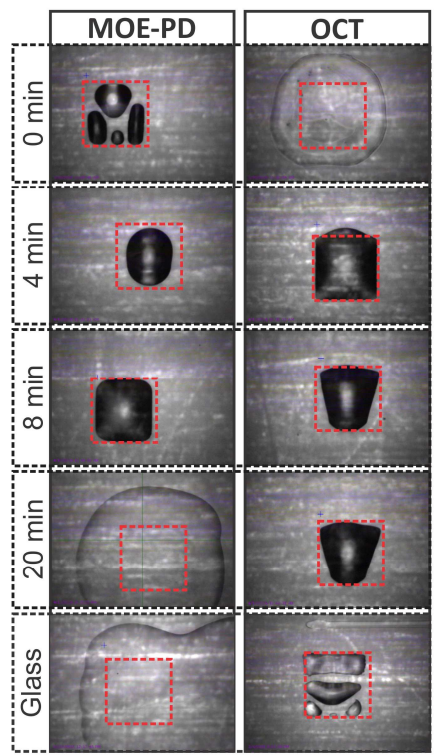

**Figure 7.** Printed square pattern of MOE–PD and OCT inks on as-deposited and partially thermally decomposed PMMA layers. The user-defined 500×500 µm$^2$ square pattern is marked in red.

**Adjusting the wetting of different substrates**

The possibility to select a substrate without concern for its wetting properties is very beneficial in printing technology. In this section we demonstrate how the wetting for the three ink–substrate systems can be adjusted using the thin polymeric layers.

The indium–tin–oxide-coated glass (ITO/glass) is used in applications that require a transparent electrode, such as photovoltaics and displays. We highlight the inkjet printing of thin dielectric structures from tantalum-oxide-based (TAS) ink.[21] Printing on bare ITO/glass results in extensive spilling of the ink, which hinders the possibility of patterning (**Figure 8**, top left). The deposition of a 10-nm-thick PMMA layer and partially decomposing it by heating at 350 °C for 8 minutes improves the wetting behavior and enables the deposition of well-defined patterns with a high resolution (**Figure 8**, top right).

The second example is printing ferroelectric lead zirconate titanate (PZT) onto a platinized silicon wafer (Pt/Si). The patterning of PZT ink is already possible on bare Pt/Si substrates; however, the printing results in some over-spilling of the ink and a large width of the printed lines (**Figure 8**, middle left). Coating the surface of the Pt/Si substrate with a 10-nm-thick PMMA layer improved the printing resolution (**Figure 8**, middle right). The increase in the contact angle is furthermore reflected in a reduction in the line width from 230 μm to 60 μm for the uncoated and the PMMA-coated Pt/Si substrate, respectively.

Another possible application for thin polymeric layers is the promotion of ink spreading on the substrate. An example is the deposition of large-area electrodes by inkjet printing commercially available silver nanoparticle ink (Ag ink) onto alumina ($Al_2O_3$) substrates. In this particular case the ink was printed with a relatively large drop spacing of 40 μm, which resulted in non-uniform layers with a broken continuity (**Figure 8**, bottom left). Coating Al$_2$O$_3$ with a hydrophilic PVP layer promoted the spreading of the ink and ensured the stable coalescence of the drops (**Figure 8**, bottom right).

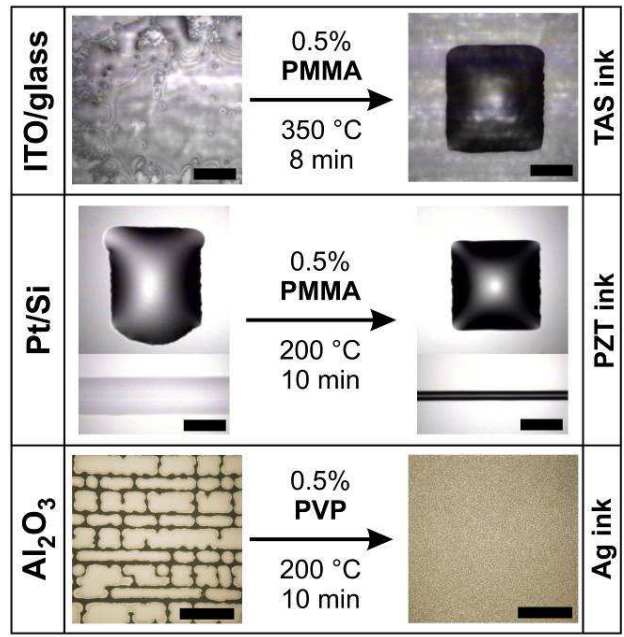

**Figure 8.** Inkjet printing of different inks on bare substrates (left) and substrates coated with thin polymeric layer (right). Scale bars correspond to 250 µm.

**Conclusions**

We described a simple and efficient approach to controlling the surface properties of different substrates using thin polymeric layers. The layers of PVA and PVP on glass exhibited hydrophilic surface properties, while the PMMA and PS layers rendered the surfaces hydrophobic. Theoretical predictions based on wetting diagrams were in good correlation with the inkjet printing of two inks with different polarities. We demonstrated that some applications, especially the printing of structures with more complex geometries, require a precise adjustment of the surface properties. The thermal decomposition of the PMMA layer produced a smooth transition from polymer- toward substrate-like surface properties, which was exploited for improving the printing outcome. The SFE analyses of the PMMA layers heated for different times at 350 °C suggest that this transition originates from the percolation of the underlying substrate surface through small, non-interacting spots in the polymer layer. Heating for different times thus provided a way to tailor the ink–substrate interactions and thus enhance the printing resolution and quality.

The practical applicability of thin polymeric layers was exemplified by three ink–substrate systems, which originally exhibited poor printing performance. Adjusting the ink–substrate interactions by implementing polymeric layers improved the wetting behavior and made it possible to print structures with a well-defined geometry and morphology.

AUTHOR INFORMATION

**Corresponding Author**

Aleksander Matavž. E-mail: aleksander.matavz@ijs.si

**Author Contributions**

The manuscript was written through contributions of all authors. All authors have given approval to the final version of the manuscript.

**Notes**

The authors declare no competing financial interest.

ACKNOWLEDGMENT

The authors acknowledge M. Širca and E. Žagar from the National Institute of Chemistry, Slovenia for providing the poly(styrene). We also acknowledge the financial support of the Slovenian Research Agency (P1-0125, P2-0105, PR 06799).

ABBREVIATIONS

PMMA, poly(methyl methacrylate), PS, poly(styrene), PVA, poly(vinyl alcohol), PVP, poly(vinyl pyrrolidone), SFE, surface free energy.

**For Table of Contents Use Only**

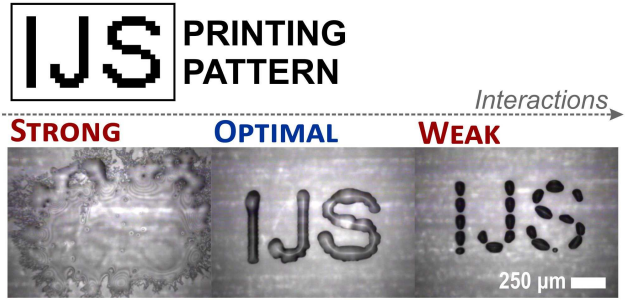